\documentclass[twocolumn,english,twocolumns]{revtex4}
\usepackage[T1]{fontenc}
\usepackage[latin9]{inputenc}
\usepackage{amsmath}
\usepackage{amssymb}
\usepackage{graphicx}
\usepackage{esint}

\makeatletter
\@ifundefined{textcolor}{}
{%
 \definecolor{BLACK}{gray}{0}
 \definecolor{WHITE}{gray}{1}
 \definecolor{RED}{rgb}{1,0,0}
 \definecolor{GREEN}{rgb}{0,1,0}
 \definecolor{BLUE}{rgb}{0,0,1}
 \definecolor{CYAN}{cmyk}{1,0,0,0}
 \definecolor{MAGENTA}{cmyk}{0,1,0,0}
 \definecolor{YELLOW}{cmyk}{0,0,1,0}
}

\usepackage{babel}

\makeatother

\usepackage{babel}
\begin{document}

\title{Measuring the degree of unitarity for any quantum process}

\author{Jing-Xin Cui  and  Z. D. Wang}

\affiliation{Department of Physics and Center of Theoretical and Computational
Physics, The University of Hong Kong, Pokfulam Road, Hong Kong, China
}

\date{\today}
\begin{abstract}
Quantum processes can be divided into two categories: unitary and non-unitary ones. 
For a given quantum process, we can define a \textit{degree of the unitarity (DU)} of this process to be the fidelity between it and its closest unitary one.
The DU, as an intrinsic property of a given quantum process, is able to quantify
the distance between the process and the group of unitary
ones, and is closely related to the noise of this quantum process.
We derive analytical results of DU for qubit unital channels, and obtain
the lower and upper bounds in general. The lower bound is tight
for most of quantum processes, and is particularly tight when the corresponding
DU  is sufficiently large. The upper bound is found
to be an indicator for the tightness of the lower bound. Moreover, we study
the distribution of DU in random quantum processes with different
environments. In particular, The relationship between the DU of any quantum process
and the non-markovian behavior of it is also addressed.
\end{abstract}

\pacs{03.67.-a, 03.65.Yz}

\maketitle

\section{Introduction\label{sec:Introduction}}

Quantum computation is of great interest in recent years for its great
speed up for solving some problems\cite{Shor1994,Grover1997}and its
efficiency in simulating physical systems\cite{Feynman1982}. Ideally,
a quantum computer is a closed quantum system, and the evolution of
the system is a unitary operation. However, as any quantum computer
is inevitably interacted with the environment, the system is more
or less open, and thus  the system evolution becomes non-unitary.
This is one of major difficulties for making a large scale quantum
computer.

When a quantum system becomes open, the information about the initial
state may be lost after the evolution of the system, and we cannot
undo a quantum operation in general. This is discussed in quantum
commutation, where quantum channel capacity is introduced to describe
the information transfer ability of a quantum channel (see, e.g.,
Ref. \cite{Nielsen2000}). Quantum channel capacity also illustrates the
quantum channel noise. Notably, there are also other ways to study the
noise of a quantum channel, such as using the addition of noise to
change a quantum map into an entanglement-breaking map~\cite{Pasquale2012}.

Since we can divide quantum operations into two categories: the unitary
operations which are ideal, and the non-unitary ones which are most
of the cases in real quantum systems, it is nature to ask to what extend a general quantum operation deviates from the
group of unitary quantum operations, which appears to be fundamentally important. We here treat all of the unitary operations
as a group because the information of a general quantum state can
be perfectly preserved for all unitary operations. 

To study the distance between a given quantum operation and the group
of unitary operations, a common way is to find the closest unitary
quantum operation for the given one, and use the distance between
these two operations as the distance between the given quantum operation
and the unitary operation group. The measure of the distance between
two quantum processes has already been well studied\cite{Nielsen,Nielsen2005,Grace2010,Zpuchala2011},
and we choose quantum process matrix fidelity in this paper. We define
the fidelity between a given quantum process and its closest unitary
one as the degree of unitarity(DU) of this given quantum process.

We think the study of the measure of the DU is important because of
the following reasons. First of all, DU quantifies the difference
between a realistic quantum operation and the ideal ones. It is an
important intrinsic property for a quantum operation. Also, the
DU of a quantum process is closely related to the noise of this process,
the quantum capacity, and some other physical quantities such as the
non-markovian behavior of a quantum process.

To obtain the DU of a given quantum process, a core problem we need
to solve is to find the closest unitary operations of the given quantum
process, since an analytical result for the measure of the distance
between two quantum processes has already been available~\cite{Nielsen,Nielsen2005,Grace2010,Zpuchala2011}.
However, finding an optimal unitary matrix is in general difficult because
the optimization should be taken over all the unitary matrices. Remarkably, we here obtain analytical results for all qubit
unital channels. Moreover, we also reveal the upper and lower bounds for
the DU of a general quantum process. The lower bound is very tight
and deviates from the true value slightly for most quantum processes,
and it can be treated as a good approximation of the DU in most cases.

Our paper is organized as follows. In the next section (Sec. II),
we give the definition of the DU of a quantum process and address
its properties. In Sec. III, we give analytical results for qubit unital
channels and some other special cases. In Sec. IV, we reveal the upper
and lower bounds for the DU of a quantum process in general cases,
and discuss the tightness of the bounds. In Sec. V, we study the probability
distribution of DU of a quantum system interacting with environment
of different dimensions. Finally, we present a brief summary and outlook in
Sec. VI.

\section{Definition and properties\label{sec:Definition-and-properties}}

A quantum process $\varepsilon$ can be expressed in Kraus operators\cite{Kraus1983},

\begin{equation}
\varepsilon(\rho)=\sum_{k}E_{k}\rho E_{k}^{\dagger},
\end{equation}

where the operators satisfy $\sum_{k}E_{k}^{\dagger}E_{k}=I$ to insure
this is a trace preserving map. The Kraus operator representation
has clear physical meaning and every part of the summation can be
treated as the evolution of the system when the environment is measured
and a specific result is obtained. But since we need a group of matrices
to represent a quantum process using Kraus operators, sometimes it
is more convenient to use quantum process matrix to describe a quantum
process which only need one matrix .

Suppose $\{|m\rangle\}$ is an orthogonal basis set for the system,
let $A_{j}=|m\rangle\langle n|$, then the quantum process $\varepsilon$
can be expressed as

\begin{equation}
\varepsilon(\rho)=\sum_{m,n}(\chi_{\varepsilon})_{mn}A_{m}\rho A_{n}^{\dagger},
\end{equation}

where $\chi$ is called the quantum process matrix which contains
all the information of the quantum process.

For a given quantum process $\varepsilon$ and a unitary operation
$U$, how to measure the distance between them has been widely discussed\cite{Nielsen,Nielsen2005,Grace2010,Zpuchala2011}.
One way is consider the difference of the out put states for $U$
and $\varepsilon$. As we want the distance between two processes
to be independent of the initial state, a nature way is to average
over all the initial states. If we choose fidelity to measure the
difference of the output states, the distance between the processes
can be measured by

\begin{equation}
F=\int d|\psi\rangle\langle\psi|U^{\dagger}\varepsilon(|\psi\rangle\langle\psi|)U|\psi\rangle.
\end{equation}

Here $|\psi\rangle$ is the initial state of the system. The integration
is carried over the whole Hilbert space of the quantum state of the
system, and the volume of the state is calculated according to Harr
measure.

The average fidelity measure not only has clear physical meaning,
but also has analytical expression. 

In Ref.\cite{Nielsen} , they give an analytical result in this formula:

\begin{equation}
F_{ave}(\varepsilon,U)=\frac{n+\sum_{k}|\text{tr}(U^{\dagger}E_{k})|^{2}}{n(n+1)},\label{eq:average fidelity}
\end{equation}

where $n$ is the dimension of the system.

Another way to measure the distance is to measure the distance between
the quantum process matrices directly. We also choose fidelity to
measure the distance between process matrices. Then the fidelity between
two processes is 

\begin{equation}
F_{pro}(\varepsilon,U)=F(\chi_{\varepsilon},\chi_{U}),
\end{equation}

where $F(\chi_{\varepsilon},\chi_{U})=\text{(tr}(\chi_{\varepsilon}^{\frac{1}{2}}\chi_{U}\chi_{\varepsilon}^{\frac{1}{2}}))^{2}$.

It turns out that $ $the average fidelity measure of two quantum
processes and the fidelity between two process matrices is related
by the formula below\cite{Nielsen2005}:

\begin{equation}
F_{ave}(\varepsilon,U)=\frac{nF_{\text{pro}}(\varepsilon,U)+1}{n+1}
\end{equation}

Substituting it into Eq. (\ref{eq:average fidelity}), we can get
that 

\begin{equation}
F_{pro}(\varepsilon,U)=\frac{\sum_{k}|\text{tr}(U^{\dagger}E_{k})|^{2}}{n^{2}}\label{eq:processfidelity}
\end{equation}

We choose $F_{\text{pro}}(\varepsilon,U)$ to measure the distance
between the quantum process $\varepsilon$ and the unitary evolution
$U$ in this paper. The reasons for this will be stated later in this
section. In the following, we denote $F_{\text{pro}}(\varepsilon,U)$
as $F(\varepsilon,U)$ for simplicity.

Now we consider the main topic of this paper, the measure of the degree
of unitarity(DU) for a given quantum process $\varepsilon$. As it
is discussed in Sec. \ref{sec:Introduction}, we choose the nearest
unitary operation $U_{0}$ for the given quantum process $\varepsilon$
and use $F(\varepsilon,U_{0})$ to measure the distance between $\varepsilon$
and $U$. We define $F(\varepsilon,U_{0})$ as the DU for $\varepsilon$. 

This definition can be treated as the geometry measure of the DU,
and the geometry measure is also used in other cases such as quantum
discord\cite{Luoshunlun2010}.

According to the definition and Eq. (\ref{eq:processfidelity}), the
DU of a quantum process $\varepsilon$ is,

\begin{equation}
DU(\epsilon)=\text{Max}\{\frac{\sum_{k}|\text{tr}(U^{\dagger}E_{k})|^{2}}{n^{2}}|U\}.\label{eq:Definition of DU}
\end{equation}

Generally, it is hard to get analytical result for the above expression
as the optimization is over all the unitary matrices. Luckily in this
problem for some special cases we can get clean result. We will discuss
this in details in the next section. 

Here we focus on the properties of DU to illustrate that this is indeed
a reasonable definition and measures the similarity between a general
quantum process and unitary ones.

First, the $DU$ of $\varepsilon$ should be a property of the process
and should not be depended on the choice of Kraus operators $E_{k}$.
This property is assured by that the average fidelity is independent
of the choice of the Kraus operators\cite{Nielsen}.

Also, the DU of $\varepsilon$ should not be smaller than the DU $f\circ\varepsilon$,
which means that the quantum process cannot become more unitary by
subsequent processes. This is also guaranteed by the properties of
average fidelity\cite{Nielsen2000}. This property also make the DU
as an indicator for non-markovian behavior of a quantum process\cite{Breuer2009}.
It can be deduced that if DU increases during a quantum evolution,
then this quantum evolution is non-markovian. However, the converse
is not true.

Next we consider the extreme values of the DU of $\epsilon$. When
$\varepsilon$ is unitary, the DU should reach its maximum. This can
be easily confirmed because in this case the nearest unitary operation
is $\varepsilon$ itself and one can get that DU of this unitary process
is 1. For the minimum of DU, intuitively, one can choose the quantum
process(or quantum channel) to be a maximal depolarizing channel,
which maps all the initial states into the maximum mixed state. In
this case, one can choose arbitrary unitary operation as the nearest
one, and the DU of this process is equal to $\frac{1}{n^{2}}$. 

Finally we discuss the difference between the DU of $\varepsilon$
and $\varepsilon\circ U$, where $U$ is a unitary operation acting
on an additional quantum system. One can verify that 

\begin{equation}
DU(\varepsilon)=DU(\varepsilon\circ U),
\end{equation}
which means that the DU of a given quantum process will not change
by adding an unitary operation imposed on an ancillary quantum systems.
Also one can verify that if we choose the average fidelity as the
measure for two quantum processes, although the nearest unitary operation
will not change, the above property will not be satisfied. This is
the reason that we choose the process matrix fidelity measure.

One can also refer to Table 1 to get a clearer picture of DU, where
the DU of some important qubit channel is listed. Table 1 can be get
using the method provided in the next section.

\section{Analytical result for some special cases}

To get the DU of a quantum process $\varepsilon$, we need to find
the nearest $U$ from the group of unitary matrices. This is in general
difficult. However, if the quantum process $\varepsilon$ satisfies
some properties, we can get analytical result. 

We introduce the inner product in the operator space, and denote the
inner product between two operators $A$ and $B$ as $\langle A,B\rangle$:

\begin{equation}
\langle A,B\rangle=\text{tr}(A^{\dagger}B)
\end{equation}

Then according to Eq. (\ref{eq:Definition of DU}), $ $the DU of
the quantum process $\varepsilon$ can be expressed as

\begin{equation}
DU(\epsilon)=\text{Max}\{\frac{\sum_{k}|\langle U,E_{k}\rangle|^{2}}{n^{2}}|U\}.
\end{equation}

So the DU of a quantum process is the sum of the square of the projections
of the nearest $U$ on each Kraus operator. If each Kraus operator
is a unitary operation multiplied by a constant, and they are orthogonal
with each other, then the optimization process can be greatly simplified.

\textit{Theorem 1:}

For a quantum process $\varepsilon(\rho)=\sum_{k}E_{k}\rho E_{k}^{\dagger}$, if $E_k=\alpha U_k$ for all $k$s and $U_k \perp U_j$ for $k \neq j$, then 
\begin{equation}
DU(\varepsilon)=|\alpha_{k_{max}}|^2,
\end{equation}
where $\alpha_{k_{max}}$ is the coeficient with the max norm.

\textit{Proof:}

Expand $\{U_{k}\}$ into a set of orthogonal basis in the operator
space with operations $\{V_{j}\}$. Then any unitary operation $U$
can be expressed as a linear combination of $\{U_{k},V_{j}\}$,

\begin{equation}
U=\sum_{k}a_{k}U_{k}+\sum_{j}b_{j}V_{j},
\end{equation}
where $\sum_{k}|a_{k}|^{2}+\sum_{j}|b_{j}|^{2}=1$.

Then 

\begin{eqnarray}
\sum_{k}|\langle U,E_{k}\rangle|^{2} & = & \sum_{k}|\langle\sum_{i}a_{i}U_{i}+\sum_{j}b_{j}V_{j}|E_{k}\rangle|^{2}\nonumber \\
 & = & \sum_{k}|\langle\sum_{i}a_{i}U_{i}|E_{k}\rangle|^{2}\nonumber \\
 & = & \sum_{k}\sum_{i}|a_{i}|^{2}\delta_{ki}|\alpha_{k}|^{2}n^{2}\nonumber \\
 & = & n^{2}\sum_{k}|a_{k}|^{2}|\alpha_{k}|^{2}
\end{eqnarray}

Suppose $\alpha_{1}$ is the coefficient with the biggest norm. As
$\sum_{k}|a_{k}|^{2}\leq1$, one can see that the above expression
reaches its maximum when $|a_{1}|=1$.

In this case, according to the expression of the DU of a quantum process
in Eq. (\ref{eq:Definition of DU})$ $,

\begin{eqnarray}
DU(\varepsilon) & = & \text{Max}\{\frac{\sum_{k}|\langle U,E_{k}\rangle|^{2}}{n^{2}}|U\}\nonumber \\
 & = & |\alpha_{1}|^{2}.
\end{eqnarray}

This completes the proof.

Of course, the situation in Theorem 1 is a very special case, it requires
all all the Kraus operators to be proportional to unitary operations
and are orthogonal to each other at the same time. But one can notice
that the representation of a quantum process using Kraus operators
has its freedom and it turns out that every quantum process can be
represented in orthogonal Kraus operators\cite{Nielsen2000}.

Suppose a quantum process $\varepsilon(\rho)=\sum_{k}E_{k}\rho E_{k}^{\dagger}$,
let $W$ be the correlation matrix with $W_{jk}=\langle E_{j},E_{k}\rangle$.
One can note that $W$ is a Hermitian matrix. We can diagonolize $W$
with unitary matrix $u$

\begin{equation}
D=uWu^{\dagger}
\end{equation}

It can be proved that the rank of $D$ is $n^{2}$ at most, where
$n$ is the dimension of the system.

Let 
\begin{equation}
F_{i}=\sum_{j}u_{ij}^{*}E_{j}
\end{equation}

One can show that $\{F_{i}\}$ is equivalent with $\{E_{k}\}$ and 

\begin{equation}
\langle F_{i},F_{k}\rangle=\delta_{ik}D_{kk}
\end{equation}

Further more, for a qubit channel, if the original Kraus operators
$E_{k}$s are proportional to unitary operations, then after diagonaliztion,
the operators $F_{k}$ are still unitary operations. This theorem
can be found in Ref.\cite{Bengtsson2006}. Here we give a different
proof using the notations in this paper.

\textit{Proof: }

Suppose the Kraus operators are $E_{k}=\sqrt{p_{k}}U_{k}$ , where
$\sum_{k}p_{k}=1$, $U_{k}^{\dagger}U_{k}=I$.

Follow the diagonalization process stated above, we can get the Kraus
operators after diagonalization $\{F_{k}\}$. Then

\begin{equation}
F_{k}^{\dagger}F_{k}=\sum_{ij}u_{kj}u_{ki}^{*}E_{j}^{\dagger}E_{i}.\label{eq:FkdaggerFk}
\end{equation}

One can note that if $E_{k}$ is replaced by $\alpha E_{k}$ with
$|\alpha|=1$, the quantum process will be the same as before. So
we can treat all the $U_{k}$s as SU(2) matrices. For a SU(2) matrix,
one can represent it as 

\begin{equation}
U=\left(\begin{array}{cc}
cos\theta e^{i\phi} & sin\theta e^{i\gamma}\\
-sin\theta e^{-i\gamma} & cos\theta e^{-i\phi}
\end{array}\right)
\end{equation}

One can verify that $\text{tr}(U)$ is a real, and 

\begin{equation}
U+U^{\dagger}=\left(\begin{array}{cc}
2cos\theta cos\phi & 0\\
0 & 2cos\theta cos\phi
\end{array}\right).
\end{equation}

Also one can notice that $W$ is an orthogonal matrix in this case,
so $u$ is real too. On the r.h.s. of Eq. (\ref{eq:FkdaggerFk}),
if $i=j$, $u_{kj}u_{ki}^{*}E_{j}^{\dagger}E_{i}$ is proportional
to $I$. If $i\neq j$, as $u$ is real and $E_{j}^{\dagger}E_{i}+E_{i}^{\dagger}E_{j}$
is proportional to $I$, $u_{kj}u_{ki}^{*}E_{j}^{\dagger}E_{i}+u_{ki}u_{kj}^{*}E_{i}^{\dagger}E_{j}$
is also proportional to $I$. So we can get

\begin{equation}
F_{k}^{\dagger}F_{k}=\alpha I,
\end{equation}
where $\alpha$ is a complex number. This condition implies that $F_{k}$
is proportional to a unitary operation.

Combined with Theorem 1 together, we can get that if a qubit channel
is a convex combination of unitary channels, we can get analytical
result for the DU of this channel.

For qubit channels, it can be shown that every unital channel is a
convex combination of unitary channels, where a unital channel is
a channel that maps $\rho=\frac{1}{n}I$ into $\rho'=\frac{1}{n}I$.
This is related to Birkhoff conjecture, which is only true for $n=2$\cite{Landau1993}.
Since we can get analytical result for convex combination of unitary
channels, we can get analytical result of the DU for all qubit unital
channels. 

Convex combination of unitary channels are of great interest as they
are the only channels that can be perfectly inverted by monitoring
the environment\cite{Gregoratti2003}. The calculation of DU for these
channels offers a new perspective for the information and the noise
of these channels.

Now we list the analytical result of DU in Table 1 for some important
qubit channels, including depolarizing channel, bit flip channel,
phase flip channel, and amplitude damping channel. We can see that
the DU of these quantum channel is between $\frac{1}{4}$ and 1, as
it is shown in Sec. \ref{sec:Definition-and-properties}. For depolarizing
channel, bit flip channel and phase flip channel, the nearest unitary
operation changes when the parameter of the channel changes. For amplitude
damping channel, the nearest unitary operation is always the identity
operation.

\begin{table*} \begin{tabular}{|c|c|c|} \hline  Channel & Kraus Operators & DU of this channel\tabularnewline \hline  \hline  depolarizing channel & $\sqrt{1-\frac{3p}{4}}\left(\begin{array}{cc} 1 & 0\\ 0 & 1 \end{array}\right)$, $\sqrt{\frac{p}{4}}\left(\begin{array}{cc} 0 & 1\\ 1 & 0 \end{array}\right)$,$\sqrt{\frac{p}{4}}\left(\begin{array}{cc} 0 & -i\\ i & 0 \end{array}\right)$,$\sqrt{\frac{p}{4}}\left(\begin{array}{cc} 1 & 0\\ 0 & -1 \end{array}\right)$ & Max\{$\frac{p}{4},1-\frac{3p}{4}$\}\tabularnewline \hline  bit flip channel & $\sqrt{p}\left(\begin{array}{cc} 1 & 0\\ 0 & 1 \end{array}\right)$,$\sqrt{1-p}\left(\begin{array}{cc} 0 & 1\\ 1 & 0 \end{array}\right)$ & Max\{$p,$$1-p$\}\tabularnewline \hline  phase flip channel & $\sqrt{p}\left(\begin{array}{cc} 1 & 0\\ 0 & 1 \end{array}\right)$,$\sqrt{1-p}\left(\begin{array}{cc} 1 & 0\\ 0 & -1 \end{array}\right)$ & Max\{$p,1-p$\}\tabularnewline \hline  amplitude damping channel & $\left(\begin{array}{cc} 1 & 0\\ 0 & \sqrt{1-\gamma} \end{array}\right)$,$\left(\begin{array}{cc} 0 & \sqrt{\gamma}\\ 0 & 0 \end{array}\right)$ & $\frac{(1+\sqrt{1-\gamma})^2}{4}$\tabularnewline \hline  \end{tabular}
\caption{The DU of some important qubit channels.}
\end{table*}

\section{Lower bound and upper bound for du in general cases\label{sec:Lower-boudn-and}}

In general cases, it is hard to get analytical result for the DU of
a quantum process. We try to get the lower bound and upper bound for
DU in this section. 

To get the lower bound, we try to find a unitary operation that is
very close to the quantum process, although it may not be the closest.
As it is hard to find the closest $U$ for all these Kraus operators,
we try to find the closest $U$ for one of them. 

One option is to find the nearest $U$ for the Kraus operator with
the largest norm after diagonalization. Suppose the Kraus operators
of $\varepsilon$ after diagonalization are $\{F_{i}\}$, and the
norm of $F_{1}$, $\sqrt{\text{tr}(F_{1}^{\dagger}F_{1})}$ , is the
largest among all the $F_{i}$s. Then the closest unitary matrix for
$F_{1}$ , which is the unitary matrix that maximize $|\text{tr}(U^{\dagger}F_{1})|$,
can be found though polar decomposition.

Suppose the polar decomposition of $F_{1}$ is

\begin{equation}
F_{1}=U_{1}\sqrt{F_{1}^{\dagger}F_{1}},
\end{equation}

Then

\begin{eqnarray}
|\text{tr}(U^{\dagger}F_{1})| & \leq & |tr(U^{\dagger}U_{1}\sqrt{F_{1}^{\dagger}F_{1}})|\nonumber \\
 & = & |\text{tr}(U^{\dagger}U_{1}\sqrt{\sqrt{F_{1}^{\dagger}F_{1}}}\sqrt{\sqrt{F_{1}^{\dagger}F_{1}}})|\nonumber \\
 & \leq & \sqrt{\text{tr}(\sqrt{F_{1}^{\dagger}F_{1}})\text{tr}(U^{\dagger}U_{1}\sqrt{F_{1}^{\dagger}F_{1}}U_{1}^{\dagger}U)}\nonumber \\
 & = & \text{tr}(\sqrt{F_{1}^{\dagger}F_{1}})\label{eq:maxinnerproduct}
\end{eqnarray}

where the equality is obtained when $U=U_{1}$. One can note that
$U_{1}$ is the nearest unitary matrix for $F_{1}$. 

We choose $U_{1}$ as the approximation of the nearest unitary operation
for $\varepsilon$, and calculate the DU of this quantum process.
Then we get the lower bound for DU of $\varepsilon$,

\begin{equation}
DU(\varepsilon)\geq\frac{\sum_{i}|\text{tr}(U_{1}^{\dagger}F_{i})|^{2}}{n^{2}}\label{eq:lowerbound1}
\end{equation}

Also as $F_{i}$s are orthogonal with each other, the contribution
of $|\text{tr}(U_{1}^{\dagger}F_{i})|^{2}$ is relatively small if
$i$ is not 1, i.e., if $F_{i}$ is not the Kraus operator with the
largest norm. So we can also choose $\frac{|\text{tr}(U_{1}^{\dagger}F_{1})|^{2}}{n^{2}}$
as a simplified lower bound, which equals $\frac{(\sum_{j}\sigma_{j})^{2}}{n^{2}}$,
where $\sigma_{j}$s are the singular value of $F_{1}$. This lower
bound equals the largest visibility the interference between $\varepsilon$
and unitary channels, and it can be got experimentally\cite{Oi2003}.

We can get the lower bound in another way. We choose the Kraus operator
that may contribute most to $DU(\varepsilon)$ as the major part,
of which the value $|\text{tr}(U^{\dagger}F_{i})|^{2}$ is the maximum.
One can note that chosen Kraus operator has the largest value $ $of
$(\sum_{j}\sigma_{j})^{2}$. $ $Note that previously we choose the
one with the largest value of $\sum_{j}\sigma_{j}^{2}$$ $. This
two standards may make different when the derivation of $\sigma_{j}$s
is large. We denote the major Kraus operator by this new standard
as $F_{0}$ , and the nearest unitary operation of $F_{0}$ as $U_{0}$,
then the new low bound for $DU(\varepsilon)$ is

\begin{equation}
DU(\varepsilon)\geq\frac{\sum_{i}|\text{tr}(U_{0}^{\dagger}F_{i})|^{2}}{n^{2}}\label{eq:lowerboud2}
\end{equation}

Now we consider the upper bound of the DU of a quantum process. We
assume the nearest $U$ of $\varepsilon$ is the nearest unitary operation
of all the Kraus operator $F_{i}$s, which means that there is a unitary
operation that makes $|\text{tr}(U^{\dagger}F_{i})|^{2}$ reaches
its largest value for all $F_{i}$s. Apparently, this is not possible
for most cases. But it give a upper bound for the DU of this quantum
process.

According to Eq. (\ref{eq:maxinnerproduct}), the largest value of
$|\text{tr}(U^{\dagger}F_{i})|^{2}$ is $ $$(\sum_{j}\sigma_{j})^{2}$.
So we can get a upper bound for $DU(\varepsilon)$,

\begin{equation}
DU(\epsilon)\leq\frac{\sum_{i}(\sum_{j}\sigma_{ij})^{2}}{n^{2}}
\end{equation}

We can show that the upper bound of $DU(\varepsilon)$ is less or
equal than 1. As 

\begin{eqnarray}
\frac{\sum_{i}(\sum_{j}\sigma_{ij})^{2}}{n^{2}} & \leq & \frac{n(\sum_{ij}\sigma_{ij}^{2})}{n^{2}}\nonumber \\
 & = & \frac{\text{tr}(\sum_{i}F_{i}^{\dagger}F_{i})}{n}\nonumber \\
 & = & 1
\end{eqnarray}

Now we consider the tightness of the lower bound and upper bound.
We denote the lower bound in Eq. (\ref{eq:lowerbound1}) as lower
bound 1, and the one in Eq. (\ref{eq:lowerboud2}) as lower bound
2.

We randomly choose a quantum process and calculate the DU of the quantum
process, two lower bounds of the DU, and the upper bound of DU. Also,
in order to study the bound behavior as a function of the DU, we choose
approximately equal amount of quantum processes for each interval
of the DU of quantum processes. As shown in Fig. 1(a), the lower bound
of DU(red stars) is a good approximation of DU(black cross) in almost
all the cases. Particularly,when DU is larger than 0.4, the lower
bound of DU(red stars)is almost the same as DU. When the DU of a quantum
process is very small, the lower bound may differs from the true value
for about 10\%. 

On the other hand, the upper bound of DU is quite loose. But it can
be shown that the upper bound may be a indicator of the behavior of
the lower bound. As shown in Fig. 1(b), the difference of two lower
bounds of DU from the true value are almost zero when the upper bound
of DU is bigger than 0.8. When the upper bound of is small, the error
of the lower bound may be large. 

\begin{figure}
\includegraphics[width=1\columnwidth,height=6cm]{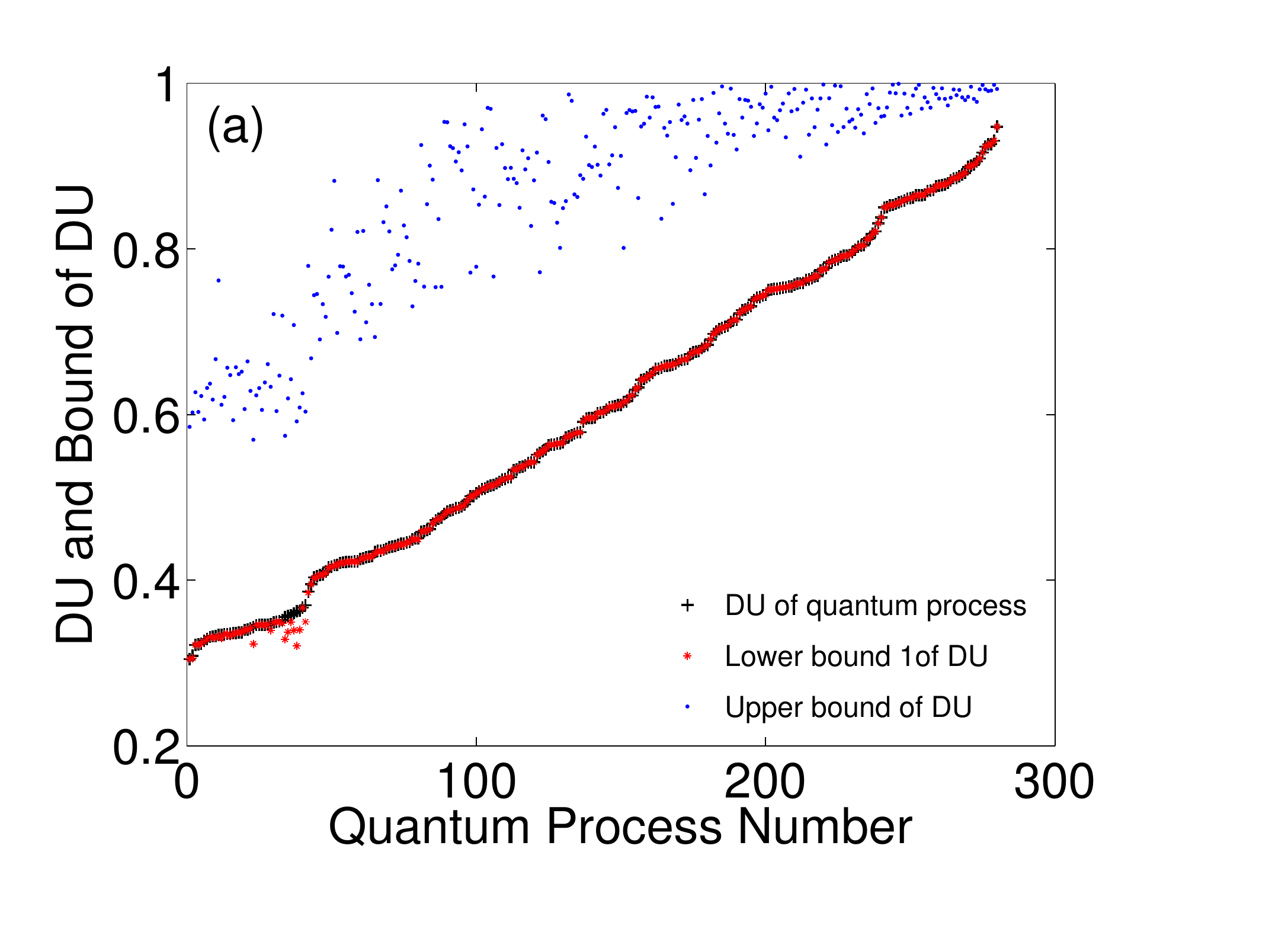}

\includegraphics[width=1\columnwidth,height=6cm]{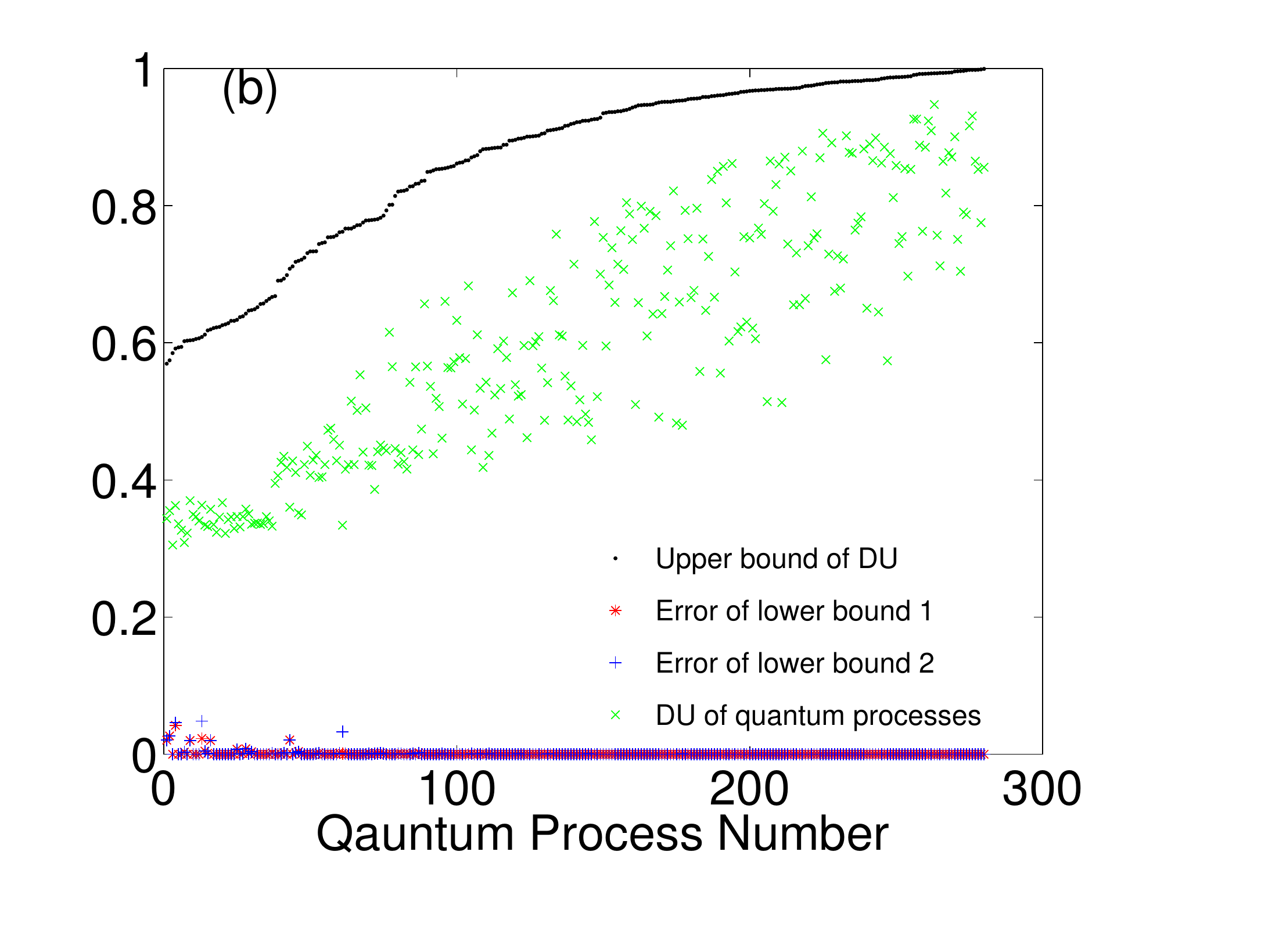}

\caption{(a)The DU of random quantum processes and the lower bound and upper
bound of the DU of the quantum process. The random quantum processes
are generated by a random unitary operation for two qubits with one
qubit traced off. The data are rearranged with the DU of them in ascending
order. (b) The error of two lower bounds of the DU of quantum processes.
The data are rearranged with the upper bound of DU in ascending order.}

\end{figure}

\section{The probability distribution of the du of a qubit system with different
environment}

In this section, we consider the probability distribution of the DU
of a random quantum process for a quantum system. This distribution
can give us some information for the DU of a random quantum process
before we really know it. It also gives the expected value of DU of
the process of a quantum system. One can note that the distribution
of DU depends on the dimension of the system and the environment.
It is also related to how the system and the environment interacts
with each other. 

Here we study a qubit system. Suppose the qubit is fully interacting
with an environment with an dimension of $d$, and the evolution of
the total system composed by the qubit and the environment is a random
unitary operation with the dimension of $2d$. In this case, we study
the probability distribution of DU of the operations imposed on the
qubit system. 

For simplicity, we only consider the cases when $d=2,4$. We generate
a random unitary operation for the total system according to the Harr
measure, and calculate the DU for the corresponding quantum operations
imposed on the principle qubit system. As it is already tested in
Sec. \ref{sec:Lower-boudn-and} that the lower bound of DU is very
tight, we treated the lower bound of DU as DU in this calculation.
As shown in Fig. 2, when the dimension of the system become bigger,
the expected value for the DU is decreased. This means that the system
is expected to be more open and the evolution of the system is expect
to be more non-unitary.

\begin{figure}
\includegraphics[width=0.5\columnwidth,height=3cm]{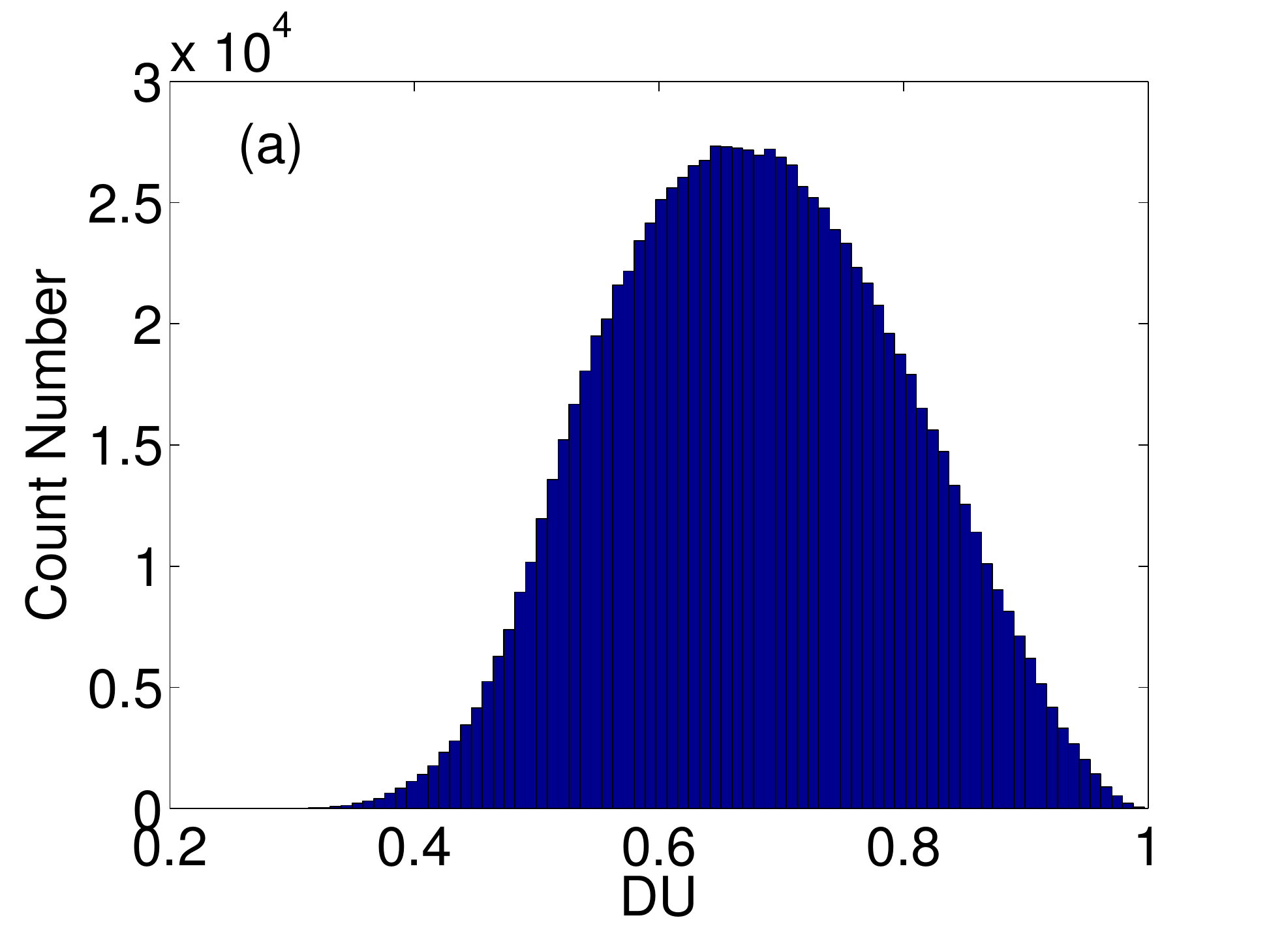}\includegraphics[width=0.5\columnwidth,height=3cm]{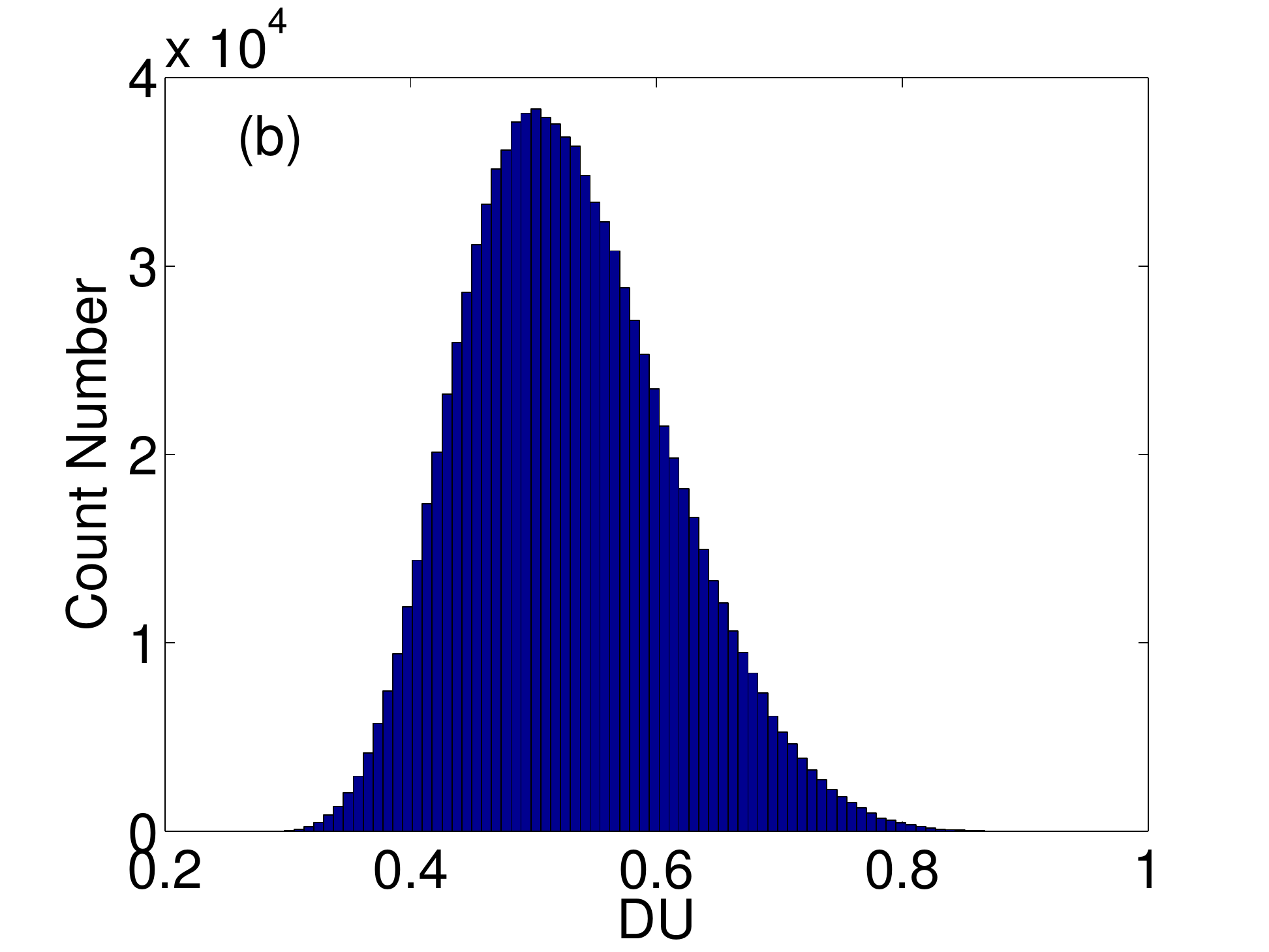}
\caption{The probability distribution of the DU of the evolution of a qubit
system interacting with an auxiliary system with the dimension of
(a) 2, (b)4. The evolution of the total system is a random unitary
operation according to Harr measure. The number of the random quantum
process is one million for each case.}
\end{figure}

\section{Summary and outlook}

In conclusion, we have investigated a key problem of the DU for any quantum
process. We have introduced a definition of the DU of a quantum process and
addressed its properties. The DU of a quantum process quantifies
the distance between a given quantum process and the group of all
unitary ones. It is closely related to the noise of the quantum
process and is an indicator for non-markovian behavior of a quantum
system.  For qubit unital channels, we have obtained an analytical result of the
DU. For general cases, we have derived the lower and upper bounds
for the DU. We have presented two different lower bounds for the DU, both being quite tight in most cases. 
The upper bound of DU can be treated
as an indicator for the tightness of DU. When the upper bound is low,
the lower bound of DU may be less tight. We have also discussed the probability
distribution of the DU of a qubit system interacting with different
environments, and found that the DU tends to become smaller when the dimension of
the environment become bigger.

The study of DU of a quantum process raises many related issues to be investigated in near future,
such as the relationship between DU of a quantum process and the noise
of a quantum process according to other measures, the DU's evolution
behavior for a real physical system, the DU for the quantum process
in an open quantum system based quantum algorithm\cite{Mizel2009,Long2011}
and so on. 
\begin{acknowledgments}

We thank Y. Hu and Z. Y. Xue for helpful discussions.
This work was supported by the RGC of Hong Kong under Grant No. HKU7058/11P.

\end{acknowledgments}

\end{document}